\documentclass[prd,twocolumn,showpacs]{revtex4}
\usepackage{mathrsfs}
\usepackage{amssymb}
\usepackage{graphicx}

\def\slashU #1{#1\kern-1.6ex\hbox{\raise0.2ex\hbox{/}}}
\def\slashL #1{#1\kern-1.2ex\hbox{\raise0.0ex\hbox{/}}}

\begin{document}
\title{Functional renormalization flow and dynamical chiral symmetry breaking of QCD}

\author{Ming-Fan Li}
\email{11006139@zju.edu.cn}
\affiliation{Zhejiang Institute of
Modern Physics, Zhejiang University, Hangzhou, PR China, 310027}

\author{Mingxing Luo}
\email{luo@zimp.zju.edu.cn}
\affiliation{Zhejiang Institute of Modern Physics, Zhejiang
University, Hangzhou, PR China, 310027}

\begin{abstract}

The dependence of functional renormalization group equation on
regulators is investigated. A parameter is introduced to control the
suppression of regulators. Functional renormalization group
equations will become regulator-independent if this newly introduced
parameter is sent to infinity in the end of calculation. One-loop
renormalization flow equations of QCD are derived. The novelty is
that both the coupling running equation and the mass running
equation are mass-dependent. Different flow patterns are explored. A
mechanism for non-occurrence of dynamical chiral symmetry breaking
is arrived at. The existence of a conformal window is also discussed
in the language of renormalization flow.

\end{abstract}

\maketitle

\section{Introduction}\label{sect:intro}

Renormalization is a concept about connecting physics at different
scales. Deriving large scale physics from small scale physics is an
important subject, and vice versa. To derive large scale
behaviors of QCD from first principle is a long-standing difficult
problem.

Functional renormalization group method originates from Wilson's
idea of integrating momentum slices successively, and studies how
theories flow between different scales. The most commonly used
formulations of functional renormalization group include
Polchinski's formalism \cite{Polchinski} and Wetterich's formalism
\cite{Wetterich} (for reviews, see for example,
\cite{Rosten-introduction,Gies-introduction}). Functional
renormalization group method uses a regulator term to modify
propagators, thereby controls contributions of modes of different
energy scales to the effective action. In the Wetterich's formalism,
for example, the regulator suppresses the low energy modes and lets
the high energy modes intact.

There have been many applications of functional remormalization
group method, for example, in the analysis of QED \cite{QED},
Non-Abelian gauge theory \cite{NonAbelian,NonAbelian-propagator},
phase transitions in gauge theories \cite{phase-transitions},
non-linear diffusion system \cite{diffusion}, Dyson's hierarchical
model \cite{Dyson-model}. For reviews of applications in specific
areas, see \cite{BCS-BEC} for BCS-BEC corssover,
\cite{quark-confinement} for quark confinement, \cite{nuclear-force}
for nuclear force, \cite{application-general} for general. Recent
applications include, sine-Gordon type models
\cite{sine-Gordon,sine-Gordon-type}, nonrelativistic inverse square
potential \cite{inverse-square}, anharmonic oscillator
\cite{anharmonic-oscillator}, Ising model \cite{Ising}, etc.

However, the use of the functional renormalization group method is
handicapped by one issue. The flow equations derived through this
method depend on regulation schemes, thus their physical meanings
are ambiguous \cite{gravity,higher-derivative,QED2}. One bypasses
this problem by arguing that flow equations become the true physical
flow equations when the cutting scale runs to zero (in the
Wetterich's formalism). But one may want not only the behavior at
zero energy scale, but also properties at finite energy scales.

In this article, we will investigate this problem closely to see
whether we can get a regulator-independent renormalization flow. The
idea is to introduce a parameter to control the suppressiveness of a
regulator. To make a complete cut, the suppressiveness below the
cut-off should be sent to infinity in the end. It is conceivable
that the result would be regulator-independent in the infinite
suppression limit, since all proper regulators then become the same:
a clear cut.

The model that we will use is QCD in $d$-dimensional Euclidian
spacetime. Recently, lattice simulations, see
\cite{AFN,DLP,AAB-etc,FHKNS,AFLNS} for example, show that there
exists a conformal window for appropriate number of fermions of QCD.
Contrary to former anticipations, dynamical chiral symmetry breaking
ceases to occur with a lower fermion number than the critical
fermion number for asymptotic freedom, $N_f^{\ddag}=11\times3/2$.
The critical fermion number for chiral symmetry breaking is accessed
to be in the interval $9<N_f^{\dag}<13$, which is slightly larger
than the lower turning point obtained from the two-loop
$\beta$-function of QCD, $51\times3/19\approx 8$.

There have been many tries to explain the existence of this
conformal window, for example, by an ansatz of all-loop
$\beta$-function \cite{all-loop-ansatz}, mass-dependent
$\beta$-functions \cite{massive-beta-function}, critical scaling
laws \cite{scaling-law}, etc.

In this article, we will tackle the problem through renormalization
flows in the $\tilde{m}^2$-$\tilde{g}^2$ plane. Here, $\tilde{m}^2$
and $\tilde{g}^2$ stand for dimensionless mass and coupling,
respectively. We will see that dynamical chiral symmetry breaking
can be ceased by and only by an IR-attractive nontrivial fixed point
with a finite dimensionless mass in the $\tilde{m}^2$-$\tilde{g}^2$
plane. All flows attracted to such a fixed point will have a
constant dimensionless mass in the infrared, which means a zero mass
in the infrared. Such a fixed point can also exist when there is a
nontrivial fixed point in the axis of the coupling of QCD. So, when
$N_f<11\times3/2$, dynamical chiral symmetry breaking may not occur
and a conformal window can open up.

The structure of this articles is as follows. In section
\ref{sec:review-of-FRG}, we briefly review the functional
renormalization group method in Wetterich formalism; in section
\ref{sec:s2-regulator}, we introduce a parameter to control the
suppressiveness of a regulator; in section
\ref{sec:application-to-qcd}, we apply the method to QCD and derive
its mass-dependent one-loop renormalization equations; in section
\ref{sec:regulator-independence}, we prove the
regulator-independence of flow equations when the suppressiveness
tends to infinity; in section \ref{sec:flow-pattern-dcsb}, we draw
flow patterns to gain insights for dynamical chiral symmetry
breaking; in section \ref{sec:conformal-window}, we show that a
conformal window can indeed turn up; finally, in section
\ref{sect:conclusion}, we give our conclusion.

\section{Brief review of the functional renormalization group}\label{sec:review-of-FRG}

Now we sketch the functional renormalization group method in the
Wetterich formalism. To start with, one adds a regulator term
$\Delta S_k$ to any given action $S$. For a scaler theory, the
regulator will have the form,
\begin{equation}\label{}
    \Delta S_k=\frac{1}{2}\int_q\varphi(-q)R_k(q)\varphi(q),
\end{equation}
where $R_k(q)=r(q^2/k^2)q^2$.
It is to suppress the low energy modes in the
momentum integration of the effective action, and keep the high energy
modes intact. Performing the integration, one is left with a
theory of low energy modes. In other words, a theory of all
range energy modes is reduced to a low energy effective theory.

For this purpose, there are some requirements on the regulator. Let
$k$ be the scale at which one wants to cut off the theory. One first
requires that $R_k(q)\sim 0$ when $q^2\gg k^2$. Secondly, one
usually takes $R_k(q)\sim k^2$ for $q^2\ll k^2$. Effectively, the
low energy modes gets a mass suppression of mass $k^2$. (Later we
will see that this requirement is not enough, which is the source of
the regulator-dependence of functional renormalization flow.) To
ensure the new theory identical to the original one when the
regulator term is removed, one needs also to have $R_k(q)\sim 0$
when $k^2\rightarrow 0$.

\begin{figure}
  \includegraphics[width=7cm]{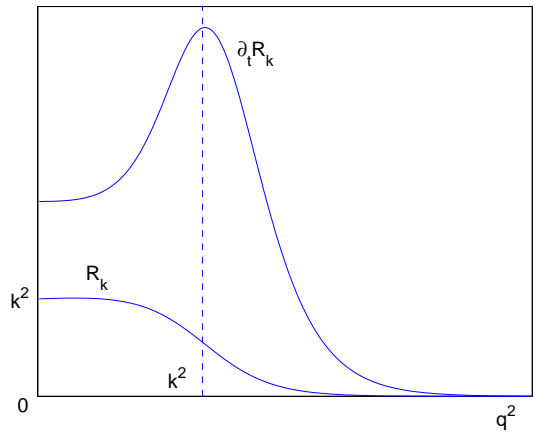}\\
  \caption{A typical profile of regulators.}\label{typical_profile}
\end{figure}

Shown in FIG. \ref{typical_profile} is the profile of a typical
regulator. Plotted also is $\partial_t R_k$
($\partial_t=k\partial_k$), which has a peak at $q^2/k^2=1$. The
peak reflects the fact that the theory is cut at $q^2=k^2$.

Regulators often used may include exponentials, such as
\begin{equation}\label{}
    r(q^2/k^2)=\frac{1}{1-e^{-q^2/k^2}}e^{-q^2/k^2},
\end{equation}
or step-functions, such as,
\begin{equation}\label{}
    r(q^2/k^2)=(k^2/q^2-1)\Theta(1-q^2/k^2),
\end{equation}
where $\Theta(x)$ is the Heaviside step function.

Let $\Gamma_k$ be the $k$-dependent effective action of the low energy
effective theory and $\tilde{\Gamma}$ be the effective
action of the regulated theory of action $S+\Delta S_k$.
The flow equation of the effective action (the Wetterich equation)
can be derived as,
\begin{equation}\label{Wetterich-equation-varphi}
    \partial_t\Gamma_k=\frac{1}{2}\text{Tr}[(\partial_tR_k)\tilde{G}_{\varphi\varphi}],
\end{equation}
where $\partial_t=k\partial_k$.
$\tilde{G}_{\varphi\varphi}=\langle\varphi\varphi\rangle-\langle\varphi\rangle\langle\varphi\rangle=(\Gamma_k^{(2)}+R_k)^{-1}$,
is the connected two-point Green function of the regulated theory.

In this equation, one observes that $\partial_t\Gamma_k$ depends
mainly on modes where $\partial_t R_k$ peaks. It reflects the fact
that the renormalization flow is driven locally and depends only on
modes in the vicinity of the cutting scale. Actually this implements
Wilson's idea of `integrating a single momentum slice'.

This equation is a functional equation of $\Gamma_k$. It is exact
and incorporates all non-perturbative effects. However, it usually
cannot be solved exactly. To extract results from this equation,
approximations must be performed. A common practice is the method of
truncation. In this approach, an ansatz of the form of the effective
action $\Gamma_k$ is made and the right hand side of the equation
(\ref{Wetterich-equation-varphi}) is projected onto the ansatz of
the effective action. Then this equation is used to derive the
flowing of coefficients in the ansatz of the effective action.

\section{Regulators}\label{sec:s2-regulator}

In this section, a parameter will be introduced to control the
suppressiveness of regulators.

As has been mentioned above, one usually requires $R_k(q)\sim k^2$
for $q^2\ll k^2$. The suppression with a mass $k^2$ may not be
enough. Instead, we would like to replace it with $R_k(q)\sim s^2
k^2$. The parameter $s^2$ controls to what degree the low energy
modes are suppressed. To completely exclude the effect of low energy
modes, we should let $s^2\rightarrow\infty$ in the end of
calculations.

Now we are to write down a typical regulator. An immediate
choice may be a step-type regulator, due to its simplicity,
\begin{equation}\label{}
    r(q^2/k^2;s^2)=(s^2k^2/q^2-1)\Theta(1-q^2/k^2).
\end{equation}
However this regulator is not continuous, and can raise problems. So
we need for continuous ones. There is a lot of smooth approximations
of the Heaviside step function $\Theta(x)$, so we can use these
functions to construct smooth regulators. For example, a
logistic-type regulator,
\begin{equation}\label{}
    r(q^2/k^2;s^2)=\frac{1}{1-e^{-q^2/(s^2k^2)}}\frac{1+e^{-s^2}}{1+e^{-s^2(1-q^2/k^2)}};
\end{equation}
an erf-type regulator,
\begin{equation}\label{}
    r(q^2/k^2;s^2)=\frac{s^2k^2}{q^2}\frac{1+\text{erf}(s^2(1-q^2/k^2))}{1+\text{erf}(s^2)};
\end{equation}
and an arctan-type regulator,
\begin{equation}\label{}
    r(q^2/k^2;s^2)=\frac{s^2k^2}{q^2}e^{-q^2/k^2}\cdot\frac{\pi/2+\text{arctan}(s^2(1-q^2/k^2))}{\pi/2+\text{arctan}(s^2)}.
\end{equation}

\begin{figure}
  \includegraphics[width=7cm]{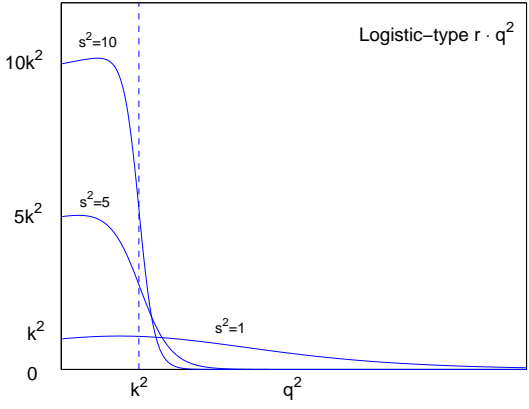}\\
  \caption{Logistic-type regulators with different values of $s^2$.}\label{logistic_profiles}
\end{figure}

In all cases, the second factor is a smooth approximation of the
step function $\Theta(1-q^2/k^2)$ which is arrived at the limit
$s^2\rightarrow\infty$. The first factor is to ensure
$r(q^2/k^2;s^2) \sim s^2k^2/q^2$ when $q^2\ll k^2$. When
$s^2\rightarrow\infty$, for all above cases, $r(q^2/k^2;s^2)\cdot
q^2$ tend to an infinity step function which takes the value of
infinity at $q^2<k^2$ and the value of zero at $q^2>k^2$. One can
see this trend in FIG. \ref{logistic_profiles}, where we have
plotted profiles of the logistic-type regulator with different
values of $s^2$.

Actually, the trend of tending to a step function and the trend of
tending to infinite suppression in the left and no suppression in
the right can always be parameterized by two independent parameters.
However, we use only one parameter $s^2$, since our later
discussions do not depend on this simplicity assumption.

To implement numerical computations, we will take a
folded-straightline regulator, in the form
\begin{equation}\label{fold-rA}
    r(q^2/k^2;s^2)\cdot q^2/k^2=\xi_0-\frac{\xi_0}{\varepsilon}\Big(\frac{q^2}{k^2}-1\Big)
\end{equation}
in the interval $[1-\varepsilon,1+\varepsilon]$, and constant
elsewhere. $s^2\rightarrow\infty$ is equivalent to
$\xi_0\rightarrow\infty$ and $\varepsilon\rightarrow 0$. The order
of taking limit is irrelevant.

\section{Application to QCD}\label{sec:application-to-qcd}

In this section we will apply the functional renormalization group
method to QCD to study its renormalization flow.


The Lagrangian of QCD in Euclidian spacetime is, \cite{Peskin}
\begin{eqnarray}
  \mathfrak{L}_E &=& \bar{\psi}^{a_f}(i\slashU{D}+m_{a_f})\psi^{a_f}+\frac{1}{4}(F^a_{\mu\nu})^2 \nonumber\\
    && ~ -\frac{1}{2\xi}(\partial^{\mu}A^a_{\mu})^2+\bar{c}^a\partial^{\mu}D^{ac}_{\mu}c^c \nonumber\\
    &=& \bar{\psi}^{a_f}(i\gamma^{\mu}\partial_{\mu}+m_{a_f})\psi^{a_f}+g\bar{\psi}^{a_f}\gamma^{\mu}t^aA^a_{\mu}\psi^{a_f}\nonumber\\
    &-&\frac{1}{2}A^a_{\mu}\partial^2A^a_{\mu}+\bar{c}^a\partial^2c^a+gf^{abc}\bar{c}^a\partial_{\mu}(A^b_{\mu}c^c) \nonumber\\
    &+&
    gf^{abc}(\partial_{\kappa}A^a_{\lambda})A^b_{\kappa}A^c_{\lambda}+\frac{1}{4}g^2f^{abe}f^{cde}A^a_{\kappa}A^b_{\lambda}A^c_{\kappa}A^d_{\lambda}.
\end{eqnarray}
We have used the convention
$\{\gamma^{\mu},\gamma^{\nu}\}=-2\delta^{\mu\nu}I_{4\times4}$ and the
gauge $\xi=1$.

Now go to momentum space and take the following ansatz for the
effective action
\begin{eqnarray}\label{action-ansats}
  \Gamma &=& \int_p Z_{\psi}\bar{\psi}^{a_f}(p)(\slashL{p}+m)\psi^{a_f}(p) \nonumber\\
         &+& \frac{1}{2}Z_A\int_p A^a_{\mu}(-p)\Big(p^2\delta^{\mu\nu}-(1-\frac{1}{\xi})p^{\mu}p^{\nu}\Big)A^a_{\nu}(p) \nonumber\\
         &-& Z_c\int_p \bar{c}^a(p)p^2c^a(p) \nonumber\\
         &+& gZ_{\psi}\sqrt{Z_A}\int_p\int_{p'}\bar{\psi}^{a_f}(p)\gamma^{\mu}t^aA^a_{\mu}(p-p')\psi^{a_f}(p')  \nonumber\\
         &+& gZ_c\sqrt{Z_A}\int_p\int_{p'}f^{abc}(ip_{\mu})\bar{c}^{a}(p)A^b_{\mu}(p-p')c^{c}(p')  \nonumber\\
         &+& gZ_A\sqrt{Z_A}\int_p\int_{p'}\int_q(2\pi)^d\delta^{(d)}(p+p'+q) \nonumber\\
         && ~~~~~~~~ \cdot f^{abc}(-iq_{\mu})A^a_{\nu}(q)A^b_{\mu}(p)A^c_{\nu}(p')  \nonumber\\
         &+& \frac{1}{4}g^2Z_A^2\int_p\int_{p'}\int_q\int_{q'}(2\pi)^d\delta^{(d)}(p+p'+q+q') \nonumber\\
         && ~~~~~~~~ \cdot f^{abe}f^{cde}A^a_{\kappa}(p)A^b_{\lambda}(p')A^c_{\kappa}(q)A^d_{\lambda}(q').
\end{eqnarray}
For simplicity, we have assumed all fermions have the same mass.


The regulators are
\begin{eqnarray}
  \Delta S_{\psi} &=& \int_p\int_q\bar{\psi}^{a_f}(p)\hat{R}_{\psi}^{a_fb_f}(p,q)\psi^{b_f}(q); \\
  \Delta S_{A} &=& \frac{1}{2}\int_p\int_q A^a_{\mu}(-p)\hat{R}_A^{ab,\mu\nu}(p,q)A^b_{\nu}(q); \\
  \Delta S_{c} &=& \int_p\int_q \bar{c}^a(p)\hat{R}^{ab}_c(p,q)c^b(q),
\end{eqnarray}
with
\begin{eqnarray}
  \hat{R}_{\psi}^{a_fb_f}(p,q) &=& Z_{\psi}\delta^{a_fb_f}\delta_{pq}r_{\psi}(p^2/k^2;s^2)\slashL{p}, \\
  \hat{R}_A^{ab,\mu\nu}(p,q) &=& Z_A\delta^{ab}\delta^{\mu\nu}\delta_{pq}r_A(p^2/k^2;s^2)p^2,\\
  \hat{R}_c^{ab}(p,q) &=& -Z_c\delta^{ab}\delta_{pq}r_A(p^2/k^2;s^2)p^2,
\end{eqnarray}
with $(1+r_{\psi})^2=(1+r_A)$.


The functional renormalization group equation can be derived as
\begin{eqnarray}\label{wetterich-eq}
    \partial_t\Gamma &=& \frac{1}{2}\text{Tr}[(\partial_t\hat{R}_A)\tilde{G}_{AA}]-\text{Tr}[(\partial_t\hat{R}_{\psi})\tilde{G}_{\psi\bar{\psi}}] \nonumber\\
                     &-& \text{Tr}[(\partial_t\hat{R}_c)\tilde{G}_{c\bar{c}}],
\end{eqnarray}
where $\tilde{G}_{AA}$, $\tilde{G}_{\psi\bar{\psi}}$ and
$\tilde{G}_{c\bar{c}}$ are connected Green functions,
\begin{eqnarray}
  \tilde{G}_{AA} &=& \langle A(q)A(p)\rangle-\langle A(q)\rangle\langle A(p)\rangle, \\
  \tilde{G}_{\psi\bar{\psi}} &=& \langle \psi(q)\bar{\psi}(p)\rangle-\langle\psi(q)\rangle\langle\bar{\psi}(p)\rangle, \\
  \tilde{G}_{c\bar{c}} &=& \langle c(q)\bar{c}(p)\rangle-\langle c(q)\rangle\langle \bar{c}(p)\rangle.
\end{eqnarray}
Since $\tilde{G}\cdot\tilde{\Gamma}^{(2)}=1$, one has
\begin{eqnarray}
  \tilde{G}_{\psi\bar{\psi}} &=& \Big[\tilde{\Gamma}^{(2)}_{\bar{\psi}\psi}-\tilde{\Gamma}^{(2)}_{\bar{\psi}A}\tilde{\Gamma}^{(2)-1}_{AA}\tilde{\Gamma}^{(2)}_{A\psi}\Big]^{-1}, \\
  \tilde{G}_{AA} &=& \Big[\tilde{\Gamma}^{(2)}_{AA}-\tilde{\Gamma}^{(2)}_{Ac}\tilde{\Gamma}^{(2)-1}_{\bar{c}c}\tilde{\Gamma}^{(2)}_{\bar{c}A}-\tilde{\Gamma}^{(2)}_{A\bar{c}}\tilde{\Gamma}^{(2)-1}_{c\bar{c}}\tilde{\Gamma}^{(2)}_{cA} \nonumber\\
                 && ~ -\tilde{\Gamma}^{(2)}_{A\psi}\tilde{\Gamma}^{(2)-1}_{\bar{\psi}\psi}\tilde{\Gamma}^{(2)}_{\bar{\psi}A}-\tilde{\Gamma}^{(2)}_{A\bar{\psi}}\tilde{\Gamma}^{(2)-1}_{\psi\bar{\psi}}\tilde{\Gamma}^{(2)}_{\psi A}\Big]^{-1}, \\
  \tilde{G}_{c\bar{c}} &=& \Big[\tilde{\Gamma}^{(2)}_{\bar{c}c}-\tilde{\Gamma}^{(2)}_{\bar{c}A}\tilde{\Gamma}^{(2)-1}_{AA}\tilde{\Gamma}^{(2)}_{Ac}\Big]^{-1}.
\end{eqnarray}


After a lengthy calculation, we arrived at the following flow
equations for the coupling $g$ and the mass $m$ (see Appendix \ref{app:Details} for the details)
\begin{eqnarray}\label{flow-of-m}
  \frac{\partial_t m}{m} &=& -\tilde{g}^2C_2(r)K_m, \\ \label{flow-of-g}
  \frac{\partial_t g}{g} &=& -\tilde{g}^2\big[C_2(r)K_1+C_2(G)K_2-N_fC(r)K_3\big],
\end{eqnarray}
Dimensionless quantities are defined as $\tilde{m}=m/k$,
$\tilde{g}^2=[\int d\Omega_d/(2\pi)^d]g^2/k^{4-d}$ and
\begin{eqnarray*}
  K_m &=& d\big[J(1,4,1;\tilde{m}^2)+J(0,2,2;\tilde{m}^2)\big] \nonumber\\
      &-& \frac{(d-1)(d-2)}{d}J(1,3,1;\tilde{m}^2); \\
  K_1 &=& (d-2)\big[J(1,4,1;\tilde{m}^2)+J(0,2,2;\tilde{m}^2)\big] \nonumber\\
      &-& \frac{4(d-2)}{d}J(-1,0,3;\tilde{m}^2)-\frac{(d-1)(d-2)}{d}J(1,3,1;\tilde{m}^2); \\
  K_2 &=& -\frac{d-2}{2}\big[J(1,4,1;\tilde{m}^2)+J(0,2,2;\tilde{m}^2)\big] \nonumber\\
      &+& \frac{2(d-2)}{d}J(-1,0,3;\tilde{m}^2) \nonumber\\
      &+& \frac{3(d-1)}{d}J(1,5,1;\tilde{m}^2)+\frac{2(d-1)}{d}J(0,3,2;\tilde{m}^2) \nonumber\\
      &-& \bigg[\frac{16(d-2)}{d(d+2)}+\frac{d-14}{2}+\frac{8}{d}\bigg]J(2,6,0;\tilde{m}^2); \\
  K_3 &=& -\frac{8}{d}J(-1,-2,3;\tilde{m}^2)+\frac{16(d+4)}{d(d+2)}J(-2,-4,4;\tilde{m}^2) \nonumber\\
      &-& \frac{64}{d(d+2)}J(-3,-6,5;\tilde{m}^2).
\end{eqnarray*}
Here $J(a,b,c;\tilde{m}^2)$ are dimensionless momentum integrals
defined by
\begin{eqnarray}\label{J-definition}
  J(a,b,c;\tilde{m}^2) \equiv \int_l\frac{(\partial_t r_A)\cdot k^{2(a+c)-d}(2\pi)^d/\int d\Omega_d}{(l^2)^a(1+r_{\psi})^b[m^2+l^2(1+r_{\psi})^2]^c}.
\end{eqnarray}

This integral encodes all regulator-relating information. It will be
investigated closely in the next section.

\section{Momentum integral and regulator-independence}\label{sec:regulator-independence}

Since the Wetterich equation has a one-loop structure, momentum
integrals appearing in flow equations all have the form of
(\ref{J-definition}), provided there is only one mass parameter in the
considered theory.

Generally the following relations hold,
\begin{eqnarray}
  && J(a,b,c;\tilde{m}^2)  \nonumber\\
  && =\tilde{m}^2J(a,b,c+1;\tilde{m}^2)+J(a-1,b-2,c+1;\tilde{m}^2).
\end{eqnarray}
\begin{eqnarray}\label{J-diff-relation}
  \frac{d J(a,b,c;\tilde{m}^2)}{d \tilde{m}^2}=-c J(a,b,c+1;\tilde{m}^2).
\end{eqnarray}

First consider the integrals with $a+c=d/2$. The integration can be
easily carried out analytically since $k\partial_k r_A=-l\partial_l
r_A$. When $\tilde{m}^2=0$ and $b/2+c-1> 0$, the result is
\begin{equation}\label{}
    J(a,b,c;0)=\frac{1}{\frac{b}{2}+c-1}.
\end{equation}
When $\tilde{m}^2\in[0,1)$ and $b/2+c-1> 0$, the result can be given
by the series,
\begin{equation}\label{J-series}
    J(a,b,c;\tilde{m}^2)=\sum_{n=0}^{\infty}\frac{(c)_n}{n!}J(a,b,c+n;0)(-\tilde{m}^2)^n,
\end{equation}
where $(c)_n$ is the rising factorial, defined as
$(c)_0=1$, $(c)_1=c$, $(c)_n=c(c+1)...(c+n-1)$. In this situation,
since the momentum integral can be carried out analytically without
any reference to the concrete form of the regulator, these results
are regulator-independent.

The integrals in the flow equations (\ref{flow-of-m}) and
(\ref{flow-of-g}) are all of this kind when $d=4$. When
$\tilde{m}^2=0$, the $K$ coefficients can be calculated as:
\begin{eqnarray*}
  K_m=3, ~~~ K_1=0, ~~~ K_2=\frac{11}{3\times2}, ~~~ K_3=\frac{4}{3\times2}.
\end{eqnarray*}
The ordinary one-loop $\beta$ function of QCD and the
ordinary mass running equation are reproduced.

The integrals with $a+c\neq d/2$ are little more complicated, though
generally the integration can not be carried out analytically. As we
will eventually take limit after integration, we can make an
approximation for the integral. As has been aforementioned, the flow
is driven only by modes in the vicinity of the cutoff scale, as
indicated by the fact that $\partial_t r_A$ peaks at $q^2/k^2=1$.
Actually the peak of the integrand is rather sharp, and values at
elsewhere are suppressed to nearly zero if $s^2$ is sufficiently
large. So we can make the approximation:
\begin{equation}\label{}
    \int_0^{\infty} d\tilde{l}^2\cdot\text{integrand} \approx \int_{1-\varepsilon}^{1+\varepsilon} d\tilde{l}^2\cdot\text{integrand}.
\end{equation}
When $s^2\rightarrow\infty$, $\varepsilon\rightarrow 0$.

The integral $J(a,b,c;0)$ can be evaluated as:
\begin{equation}\label{}
    \int_{\tilde{l}^2=1-\varepsilon}^{\tilde{l}^2=1+\varepsilon}l^{d-1}dl\frac{\partial_t r_A\cdot k^{2(a+c)-d}}{(l^2)^{a+c}(1+r_A)^{b/2+c}}.
\end{equation}
When $a+c-d/2=0$ and $b/2+c-1>0$, this integral can be carried out
directly, yielding $1/(b/2+c-1)$ if $s^2\rightarrow \infty$ in the
end, which is the same result as derived formerly. When $a+c-d/2\neq
0$ and $b/2+c-1>0$, this integral can also be done through the mean
value theorem of integration, yielding the same result $1/(b/2+c-1)$
if $s^2\rightarrow \infty$ in the end. These results are both
independent of the specific form of the regulator $r_A(q^2/k^2;s^2)$,
hence regulator-independent.

\section{Flow patterns and chiral symmetry
breaking}\label{sec:flow-pattern-dcsb}

In this section, we first explore different kinds of flow patterns.
A mechanism for the restoration of chiral symmetry will be arrived
at. Then we prove this mechanism indeed applies to QCD. In this
section the discussions will be constrained to situations of $d=4$.

\begin{figure}
  \includegraphics[width=8cm]{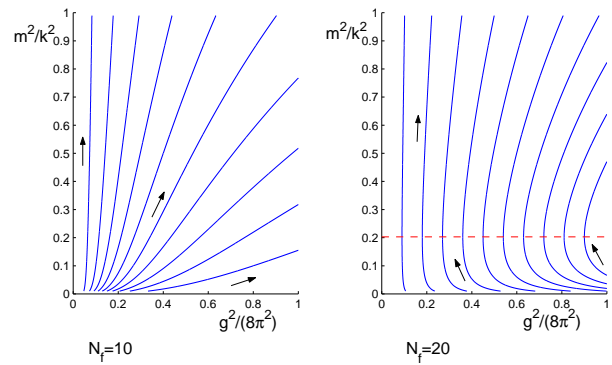}\\
  \caption{Flow patterns with $d=4$ and complete suppression. The arrows point to infrared. The red dashed line stands for $\beta=0$.}\label{FIG:flow_d_4}
\end{figure}

First consider the flow patterns of the one-loop renormalization
group of QCD, (\ref{flow-of-m}) and (\ref{flow-of-g}). The
coefficient of $\tilde{g}^2$ in the equation (\ref{flow-of-g})
always has a zero only if $N_f>11\times 3/2$. While $K_m$ is always
positive, so there is no non-trivial fixed-point for the flow
equations (\ref{flow-of-m}) and (\ref{flow-of-g}). Shown in FIG.
\ref{FIG:flow_d_4} are their flow patterns. The left panel is
typical for $N_f<11\times 3/2$, and the right panel is typical for
$N_f>11\times 3/2$.

These flow patterns do not capture the non-occurrence of dynamical
chiral symmetry breaking. However there are other kinds of flow
patterns that may do so. To explore these ones, it is natural to add
in two loop contributions as the next step.

Let's recall the ordinary two-loop renormalization group equation
for the mass \cite{two-loop-mass},
\begin{eqnarray}\label{massless-two-loop-mass}
  \partial_t\tilde{m}^2 &=& -2\tilde{m}^2(1+h_1\tilde{g}^2+h_2(\tilde{g}^2)^2),
\end{eqnarray}
with
\begin{eqnarray}\label{h-1}
  h_1 &=& 3C_2(r), \\ \label{h-2}
  h_2 &=& \frac{3}{4}C_2(r)^2+\frac{97}{12}C_2(r)C_2(G)-\frac{5}{6}C_2(r)N_f.
\end{eqnarray}
Observe that, in $h_2$ the coefficient before $N_f$ is negative. So
for large enough $N_f$, this term will dominate and the flow
equation of the mass will develop a zero point. In turn, the
$\tilde{m}^2$-$\tilde{g}^2$ flows will develop a nontrivial fixed
point. However, coefficients in this equation are independent of
$\tilde{m}^2$. In other words, this equation corresponds to Eq.
(\ref{flow-of-m}) with $\tilde{m}^2=0$ but at two-loop. If we use it
directly to depict the flows, the fixed point will be displaced and
streamlines distorted. Even worse, the flow patterns may be changed.
Nevertheless, to see how flow patterns capture the non-occurrence of
dynamical chiral symmetry breaking, it will be enough to make a
parametrization of Eq. (\ref{massless-two-loop-mass}).

Literally, we multiply two-loop terms in Eq.
(\ref{massless-two-loop-mass}) by factors
$\beta_i^{\alpha_i}/(\tilde{m}^2+\beta_i)^{\alpha_i}$ ($\alpha_i$
and $\beta_i$ are constant parameters), and add them to Eq.
(\ref{flow-of-m}). This form of mass dependence is motivated by the
$J$-integral (\ref{J-definition}). $\alpha_i$ is roughly the number
of fermion propagators in two loop momentum integrals. The
difference among $\beta_i$ can be generated after loop momentum
integration.

The resulting equation of these manipulations will be used to
explore different flow patterns. One can see that the resulting
equation may not be the true one of QCD, but it is enough for
qualitative investigation.

Concretely, the following system will be considered:
\begin{eqnarray}\label{flow-of-m-two-loop}
  \frac{\partial_t\tilde{m}^2}{2\tilde{m}^2} &=& -\Big[1+\tilde{g}^2C_2(r)K_m+H_2(\tilde{m}^2)(\tilde{g}^2)^2\Big], \\ \label{flow-of-g-4}
  \frac{\partial_t\tilde{g}^2}{2\tilde{g}^2} &=& -\tilde{g}^2\big[C_2(r)K_1+C_2(G)K_2-N_fC(r)K_3\big],
\end{eqnarray}
with
\begin{equation}\label{}
    H_2(\tilde{m}^2)=h_{21}\rho_1^{\alpha_1}+h_{22}\rho_2^{\alpha_2}-h_{23}\rho_3^{\alpha_3},
\end{equation}
where
\begin{eqnarray}
  h_{21} &=& \frac{97}{12}C_2(r)C_2(G), \\
  h_{22} &=& \frac{3}{4}C_2(r)^2, \\
  h_{23} &=& \frac{5}{6}C_2(r)N_f,
\end{eqnarray}
\begin{equation}
    \rho_i=\frac{\beta_i}{\tilde{m}^2+\beta_i}, ~~ i=1,2,3.
\end{equation}

This system is not the one of QCD. However, since our purpose is to
explore different kinds of flow patterns rather than to discuss QCD
directly, it is reasonable to use such a system. Actually, to
explore different kinds of flow patterns, any functions are
permissible in the right hand of the flow equations
(\ref{flow-of-m-two-loop}) and (\ref{flow-of-g-4}). The only problem
is whether they can lead to new and meaningful flow patterns.

One loop flow equations of QCD, (\ref{flow-of-m}) and
(\ref{flow-of-g}), do not capture the restoration of chiral symmetry,
but they provide a base to be perturbed
and to generate new flow equations.

Besides the trivial fixed point $(\tilde{m}^2=0,\tilde{g}^2=0)$, the
flow equations (\ref{flow-of-m-two-loop}) and (\ref{flow-of-g-4})
now may have a nontrivial fixed point. Taking $\tilde{m}^2$ as
the argument, the possible nontrivial fixed point is
\begin{eqnarray}
  N_f^* &=& \frac{C_2(r)K_1+C_2(G)K_2}{C(r)K_3}, \\
  \tilde{g}^{2*} &=& \frac{-C_2(r)K_m-\sqrt{[C_2(r)K_m]^2-4H_2}}{2H_2}.
\end{eqnarray}
Another unphysical root has been discarded. To ensure
$\tilde{g}^{2*}>0$, the sufficient and necessary condition is:
\begin{equation}\label{}
    H_2(\tilde{m}^{2*})<0.
\end{equation}

To see the behavior of flows near the above fixed point,
we estimate the critical exponents. If $\lambda_1$ and
$\lambda_2$ are the critical exponents of this fixed point, then,
\begin{eqnarray*}
  \lambda_1\lambda_2 &\approx& \Big[-1/\tilde{g}^{2*}+H_2(\tilde{m}^{2*})\Big]\bigg[\tilde{g}^{2*}N_fC(r)\frac{dK_3}{d\tilde{m}^2}\bigg|_{\tilde{m}^{2*}}\bigg]>0;
\end{eqnarray*}
\begin{eqnarray*}
  \lambda_1+\lambda_2 &=& -2\tilde{m}^{2*}\bigg[\tilde{g}^{2*}C_2(r)\frac{d K_m}{d\tilde{m}^2}\bigg|_{\tilde{m}^{2*}}+\frac{dH_2}{d\tilde{m}^2}\bigg|_{\tilde{m}^{2*}}\cdot(\tilde{g}^{2*})^2\bigg] \\
                      &\approx& -2\tilde{m}^{2*}\frac{dH_2}{d\tilde{m}^2}\bigg|_{\tilde{m}^{2*}}\cdot(\tilde{g}^{2*})^2 \\
                      &\varpropto& -\frac{dH_2}{d\tilde{m}^2}\bigg|_{\tilde{m}^{2*}}.
\end{eqnarray*}
When
\begin{equation}\label{}
    \frac{dH_2}{d\tilde{m}^2}\bigg|_{\tilde{m}^{2*}}<0 ~ (>0),
\end{equation}
the fixed point is IR-attractive (IR-repulsive).

\begin{figure}
  \includegraphics[width=7.5cm]{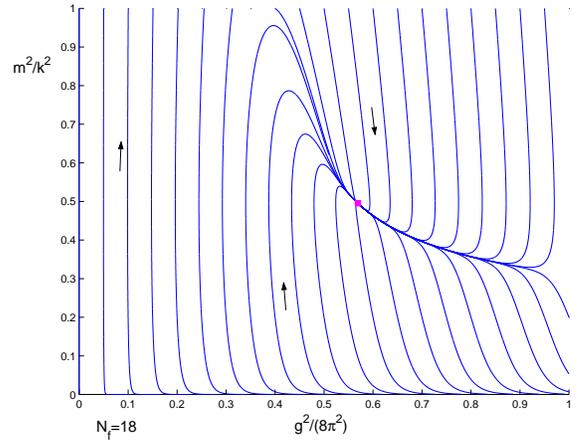}\\
  \caption{IR-attractive nontrivial fixed point. The arrows point to infrared. $(\alpha_i,\beta_i)=(3,4,5,1,1,36)$.}\label{flow_4_18}
\end{figure}

\begin{figure}
  \includegraphics[width=7.5cm]{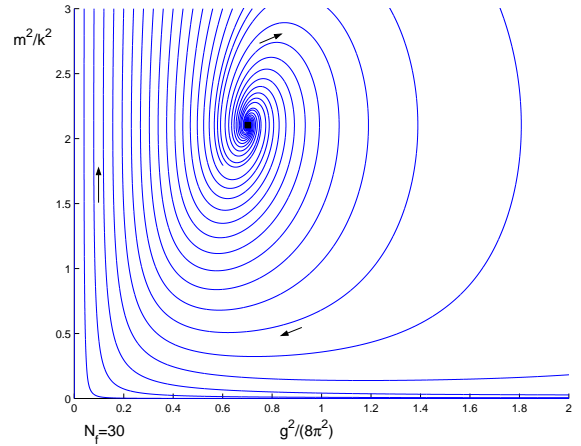}\\
  \caption{IR-replusive nontrivial fixed point. The arrows point to infrared. $(\alpha_i,\beta_i)=(2,3,4,1,2,5)$.}\label{flow_4_30_UV}
\end{figure}

One sees that the fixed-point structure of Eqs.
(\ref{flow-of-m-two-loop}) and (\ref{flow-of-g-4}) depends on the
choice of parameters $(\alpha_i,\beta_i)$. For example, the choice
of
$(\alpha_1,\alpha_2,\alpha_3,\beta_1,\beta_2,\beta_3)=(3,4,5,1,1,36)$
will lead to an IR-attractive nontrivial fixed point for all
$N_f>17.12$. While the choice of $(2,3,4,1,2,5)$ will result in an
IR-repulsive fixed point for all $N_f>25.56$. We have depicted some
flow patterns in FIG. \ref{flow_4_18} - FIG. \ref{flow_4_30_UV}.
\begin{figure}
  \includegraphics[width=7.5cm]{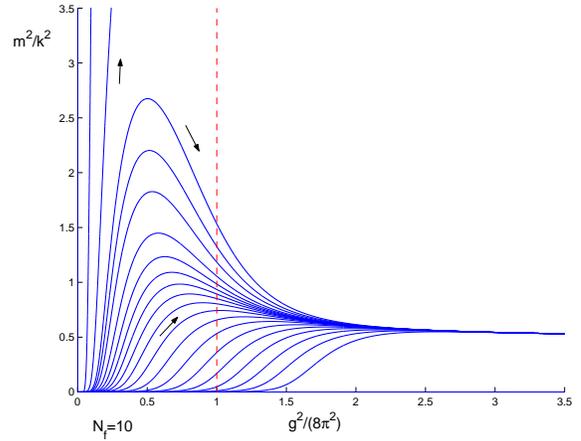}\\
  \caption{No nontrivial fixed point. The arrows point to infrared. $(\alpha_i,\beta_i)=(3,4,5,1,1,36)$.}\label{flow_4_10}
\end{figure}

Shown in FIG. \ref{flow_4_18} is the flow pattern for
$N_f=18$ with $(\alpha_i,\beta_i)=(3,4,5,1,1,36)$.
One sees that all flows are going to the IR-attractive fixed point. At this point,
$\tilde{m}^2$ is a constant, which means $m^2=\tilde{m}^2k^2$ goes
to zero when $k^2\rightarrow 0$. So there is no mass generation or
chiral symmetry breaking.

Dynamical chiral symmetry breaking is in principle a non-perurbative
behavior, much the same as color confinement. However, color
confinement can be gauged via perturbative calculation, in the form
of asymptotic freedom. Similarly, dynamical chiral symmetry breaking
may also be gauged via perturbative calculation. To chiral symmetry
the fermion mass is more relevant than the gauge coupling. So the
effect of fermion mass seems more important than higher order
quantum corrections. Thus the above discussion from the perspective
of perturbative renormaliztion flow is meaningful.

Shown in FIG. \ref{flow_4_10} is the flow pattern for $N_f=10$ with
$(\alpha_i,\beta_i)=(3,4,5,1,1,36)$. There is no nontrivial fixed
point and all flows to the limit $(\infty,\tilde{m}^2(H_2=0))$. This
cannot be interpreted as a restoration of chiral symmetry, since the coupling $\tilde{g}^2$ has gone beyond the
perturbative region. Shown in FIG. \ref{flow_4_30_UV} is the flow
pattern for $N_f=30$ with $(\alpha_i,\beta_i)=(2,3,4,1,2,5)$. An
IR-repulsive nontrivial fixed point turns up.

In the above, different flow patterns are displayed. One of them, in
FIG. \ref{flow_4_18}, provides a mechanism to explain the
non-occurrence of dynamical chiral symmetry breaking. Theories
attracted to an IR-attractive nontrivial fixed point of a finite
$\tilde{m}^{2*}$ will show no dynamical chiral symmetry breaking.

Now we show this mechanism can also apply to QCD. We shall prove the
inverse proposition: a non-occurrence of dynamical chiral symmetry
breaking indicates the theory is attracted to an IR-attractive
nontrivial fixed point of a finite $\tilde{m}^{2*}$.

The mass running equation of all-loop QCD can be written as
\begin{eqnarray}\nonumber
  \partial_t\tilde{m}^2 &=& -2\tilde{m}^2\Big[1+\sum_{i=1}^{\infty}H_i(\tilde{m}^2;N_f)(\tilde{g}^2)^i\Big].
\end{eqnarray}
Here $H_i(\tilde{m}^2;N_f)$ is the mass-dependent coefficient of the
$i$-loop correction to the fermion mass in QCD. Restoration of
chiral symmetry requires $m^2=0$ in the infrared. So the scaling
exponent of $\tilde{m}^2$ is smaller than 2, or the value of sum of
terms in the square bracket is smaller than 1. This is possible,
because when $N_f$ large enough, some of $H_i(\tilde{m}^2;N_f)$ will
be negative. Since $H_i(\tilde{m}^2;N_f)$ are loop corrections to
the fermion mass, they will all tend to 0 when
$\tilde{m}^2\rightarrow\infty$ because of the fermion propagators in
the loop integrals. Actually when $\tilde{m}^2$ large enough and
$\tilde{g}^2$ small enough, the sum of quantum corrections will be
positive. This is due to the fact: loop integrals with more than one
fermion propagators then will sub-dominate, and the one-loop
contribution will dominate.

This discussion can also apply to the non-perturbative situation.
Loop integrals with more than one fermion propagators sub-dominate
only if $\tilde{m}^2$ is sufficiently large. The difference is that
the one-loop integral is now with a gluon propagator that has been
renormalized in pure gauge theory. It is still positive, for a gluon
cannot change to a ghost by mere renormalization. One can see that
the above discussion is not based on perturbative expansion, but on
$1/\tilde{m}^2$ expansion.

So, if there is no dynamical chiral symmetry breaking, $\tilde{m}^2$
should be a finite value in the infrared. The coupling $\tilde{g}^2$
should be a nonzero (finite) value as well. The theory is attracted
to an IR-attractive nontrivial fixed point of a finite
$\tilde{m}^{2*}$. Thus we proved the proposition.

\section{The conformal window}\label{sec:conformal-window}

In this section, we will see how conformal window is reflected in
renormalization flows.

Recall the ordinary two-loop running equation of the
coupling in QCD:
\begin{equation}\label{flow-of-g-two-loop}
  \partial_t\tilde{g}^2=-\tilde{g}^2(f_1\tilde{g}^2+f_2(\tilde{g}^2)^2),
\end{equation}
with
\begin{eqnarray}
  f_1 &=& \frac{11}{3}C_2(G)-\frac{4}{3}C(r)N_f, \\
  f_2 &=& \frac{17}{3}C_2(G)^2-\frac{10}{3}C_2(G)C(r)N_f-2C_2(r)C(r)N_f.
\end{eqnarray}
One can do similar manipulations as those for the mass running
equation, however there is another choice which is simpler and more
certain.

In Eq. (\ref{flow-of-g-two-loop}), there
are two turning points for $N_f$ at which the behavior of
$\beta_{\tilde{g}^2}(=\partial_t\tilde{g}^2)$ changes.
$\beta_{\tilde{g}^2}$ is always positive for $N_f>11\times3/2$
and always negative for $N_f<51\times3/19$.
When
$51\times3/19<N_f<11\times3/2$, the sign of $\beta_{\tilde{g}^2}$
changes at a nontrivial fixed point, $\tilde{g}^{2*}=-f_1/f_2$.

This property can also be found
in the one-loop $\beta$-function of $d$-dimensional QCD with
$d=4-\epsilon<4$:
\begin{equation}
  \partial_t\tilde{g}^2=-\tilde{g}^2(\epsilon +\tilde{g}^2F_1(0)),
\end{equation}
where $F_1(0)$ is the value of $2[C_2(r)K_1+C_2(G)K_2-N_fC(r)K_3]$
with $\tilde{m}^2=0$. When $N_f<[C_2(r)K_1+C_2(G)K_2]/[C(r)K_3]$ at
$\tilde{m}^2=0$, $\beta_{\tilde{g}^2}$ is always negative. When
$N_f>[C_2(r)K_1+C_2(G)K_2]/[C(r)K_3]$ at $\tilde{m}^2=0$, the sign
of $\beta_{\tilde{g}^2}$ changes at a nontrivial fixed point,
$\tilde{g}^{2*}=-\epsilon/F_1(0)$. The difference is that
$[C_2(r)K_1+C_2(G)K_2]/[C(r)K_3]$ now plays the role of the lower
turning point, while the larger turning point now goes to infinity.

\begin{figure}
  \includegraphics[width=7.5cm]{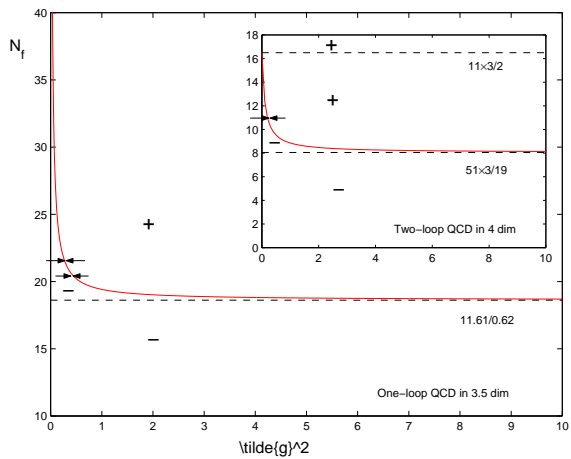}\\
  \caption{The similarity between the two situations.}\label{FIG:massless_fixedpoint}
\end{figure}

One can see this similarity from FIG. \ref{FIG:massless_fixedpoint}.
So we can use the one-loop $\beta$-function of $d$-dimensional QCD
to mimic the two-loop $\beta$-function of 4-dimensional QCD. The
following system will be considered:
\begin{eqnarray}
  \partial_t\tilde{m}^2 &=& -2\tilde{m}^2\Big[1+\tilde{g}^2C_2(r)K_m+H_2(\tilde{m}^2)(\tilde{g}^2)^2\Big], \\
  \partial_t\tilde{g}^2 &=& -2(\tilde{g}^2)^2\big[C_2(r)K_1+C_2(G)K_2-N_fC(r)K_3\big] \nonumber\\
                        && -\tilde{g}^2\epsilon.
\end{eqnarray}
This system deviates slightly from the one of the last section, so
the flows will not be changed drastically. Since now there is a
nontrivial fixed point on the axis of the coupling, there would be
modifications.
\begin{figure}
  \includegraphics[width=7.5cm]{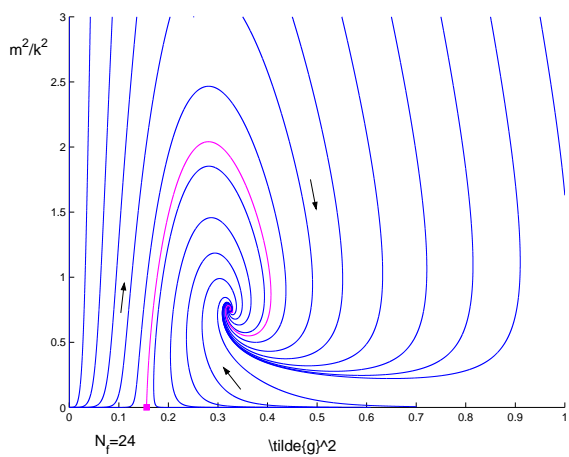}\\
  \caption{IR-attractive nontrivial fixed points. The arrows point to infrared. $(\alpha_i,\beta_i)=(3,4,5,1,1,36)$.}\label{flow_35_24}
\end{figure}

Shown in FIG. \ref{flow_35_24} is the flow pattern for the situation
of $d=3.5$, $N_f=24$ and $(\alpha_i,\beta_i)=(3,4,5,1,1,36)$. The
pattern is much the same as that of FIG. \ref{flow_4_18} except for
the fixed point on the axis. All flows are going to the nontrivial
fixed point with a finite dimensionless mass, which means there is
still no chiral symmetry breaking. Mapped onto two-loop $d=4$ QCD,
there is no chiral symmetry breaking when $N_f<11\times3/2$ and
larger than some critical number. So a conformal window opens up.

\section{Conclusion}\label{sect:conclusion}

In conclusion, the functional renormalization group method was
applied to QCD. There are variations in literature about application
of functional renormalization group method to theories with both
fermions and bosons. Hopefully our calculation can serve to help
clarify.

A parameter was introduced to discuss the dependence of flow
equations on regulators and it was shown that flow equations will be
independent of regulators if this parameter is sent to infinity in
the end of calculation.

Different flow patterns in the $\tilde{m}^2$-$\tilde{g}^2$ plane
were explored. (By no means, all possible ones were reached.) One of
them provides a mechanism to explain non-occurrence of dynamical
chiral symmetry breaking. However this flow pattern was obtained
from flow equations that were not of QCD. To show the relevance of
this mechanism to QCD, it was proved that it indeed applies to QCD,
in terms of an $1/\tilde{m}^2$ expansion of the fermion mass
renormalization group equation of QCD. One saw that dynamical chiral
symmetry breaking cannot occur if and only if the theory is
attracted to an IR-attractive non-trivial fixed point of a finite
dimensionless mass.

The existence of the conformal window was also discussed in the
language of renormalization flow.

\section*{Acknowledgement}

The work is supported in part by the National Science Foundation of
China (10875103, 11135006), National Basic Research Program of China
(2010CB833000).

\begin{appendix}

\section{Details}\label{app:Details}

\subsection{Preparation}

If we label the fields as $(\bar{\psi},\psi,A,\bar{c},c)$, and let
$(J_{\bar{\psi}},J_{\psi},J_A,J_{\bar{c}},J_c)$ be their sources,
and $\tilde{W}$ be the generation functional of connected Green
functions, the two-point connected Green function has the
structure:
\begin{equation}\label{}
    \tilde{G}=\left(
  \begin{array}{ccccc}
    ... & \tilde{G}_{\bar{\psi}\psi} & \tilde{G}_{\bar{\psi}A} & ... & ... \\
    \tilde{G}_{\psi\bar{\psi}} & ... & \tilde{G}_{\psi A} & ... & ... \\
    \tilde{G}_{A\bar{\psi}} & \tilde{G}_{A\psi} & \tilde{G}_{AA} & \tilde{G}_{A\bar{c}} & \tilde{G}_{Ac} \\
    ... & ... & \tilde{G}_{\bar{c}A} & ... & \tilde{G}_{\bar{c}c} \\
    ... & ... & \tilde{G}_{c A} & \tilde{G}_{c\bar{c}} & ... \\
  \end{array}
\right).
\end{equation}
The terms represented by ellipsis will not be used. Other terms are
defined by
\begin{eqnarray}
  \tilde{G}_{\bar{\psi}\psi} \equiv \frac{\delta^2\tilde{W}}{\delta_LJ_{\bar{\psi}}\delta_RJ_{\psi}}, ~~  \tilde{G}_{\bar{\psi}A} \equiv \frac{\delta^2\tilde{W}}{\delta_LJ_{\bar{\psi}}\delta J_A}, ...
\end{eqnarray}

The $\Gamma$ Hessian matrix has the structure,
\begin{equation}\label{}
    \Gamma^{(2)}=\left(
  \begin{array}{ccccc}
    0 & \Gamma^{(2)}_{\bar{\psi}\psi} & \Gamma^{(2)}_{\bar{\psi}A} & 0 & 0 \\
    \Gamma^{(2)}_{\psi\bar{\psi}} & 0 & \Gamma^{(2)}_{\psi A} & 0 & 0 \\
    \Gamma^{(2)}_{A\bar{\psi}} & \Gamma^{(2)}_{A\psi} & \Gamma^{(2)}_{AA} & \Gamma^{(2)}_{A\bar{c}} & \Gamma^{(2)}_{Ac} \\
    0 & 0 & \Gamma^{(2)}_{\bar{c}A} & 0 & \Gamma^{(2)}_{\bar{c}c} \\
    0 & 0 & \Gamma^{(2)}_{c A} & \Gamma^{(2)}_{c\bar{c}} & 0 \\
  \end{array}
\right).
\end{equation}

Elements of the $\Gamma$ Hessian matrix are,
\begin{eqnarray}
  \Gamma^{(2)}_{\bar{c}c} &\equiv& \frac{\delta^2\Gamma}{\delta_L\bar{c}^a(l)\delta_Rc^b(l')} \nonumber\\
                          &=& -Z_c\delta^{ab}\delta_{ll'}l^2-gZ_c\sqrt{Z_A}f^{abc}(il_{\kappa})A^c_{\kappa}(l-l'); \nonumber
\end{eqnarray}
\begin{eqnarray}
  \Gamma^{(2)}_{c\bar{c}} &\equiv& \frac{\delta^2\Gamma}{\delta_Lc^b(l')\delta_R\bar{c}^a(l)} \nonumber\\
                          &=& Z_c\delta^{ab}\delta_{ll'}l^2+gZ_c\sqrt{Z_A}f^{abc}(il_{\kappa})A^c_{\kappa}(l-l'); \nonumber
\end{eqnarray}
\begin{eqnarray}
  \Gamma^{(2)}_{\bar{c}A} \equiv \frac{\delta^2\Gamma}{\delta_L\bar{c}^a(l)\delta
  A^b_{\nu}(l')}=gZ_c\sqrt{Z_A}f^{abc}(il_{\nu})c^c(l-l'); \nonumber
\end{eqnarray}
\begin{eqnarray}
  \Gamma^{(2)}_{A\bar{c}} \equiv \frac{\delta^2\Gamma}{\delta A^b_{\nu}(-l')\delta_R\bar{c}^a(l)}=-gZ_c\sqrt{Z_A}f^{abc}(il_{\nu})c^c(l+l'); \nonumber
\end{eqnarray}
\begin{eqnarray}
  \Gamma^{(2)}_{Ac} &\equiv& \frac{\delta^2\Gamma}{\delta A^a_{\mu}(-l)\delta_R c^b(l')} \nonumber\\
                    &=&
                    gZ_c\sqrt{Z_A}f^{cab}i(-l_{\mu}+l'_{\mu})\bar{c}^c(-l+l'); \nonumber
\end{eqnarray}
\begin{eqnarray}
  \Gamma^{(2)}_{cA} &\equiv& \frac{\delta^2\Gamma}{\delta_L c^b(l')\delta A^a_{\mu}(l)} \nonumber\\
                    &=&
                    -gZ_c\sqrt{Z_A}f^{cab}i(l_{\mu}+l'_{\mu})\bar{c}^c(l+l'); \nonumber
\end{eqnarray}

\begin{eqnarray}
  \Gamma^{(2)}_{\bar{\psi}\psi} &\equiv& \frac{\delta^2\Gamma}{\delta_L\bar{\psi}^{a_f}(l)\delta_R\psi^{b_f}(l')} \nonumber\\
                                &=&
                                Z_{\psi}\delta^{a_fb_f}\delta_{ll'}(\slashL{l}+m)+gZ_{\psi}\sqrt{Z_A}\delta^{a_fb_f}\gamma^{\mu}t^aA^a_{\mu}(l-l'); \nonumber
\end{eqnarray}
\begin{eqnarray}
  \Gamma^{(2)}_{\psi\bar{\psi}} &\equiv& \frac{\delta^2\Gamma}{\delta_L\psi^{b_f}(l')\delta_R\bar{\psi}^{a_f}(l)} \nonumber\\
                                &=&
                                -Z_{\psi}\delta^{a_fb_f}\delta_{ll'}(\slashL{l}^T+m)-gZ_{\psi}\sqrt{Z_A}\delta^{a_fb_f}[\gamma^{\mu}t^a]^TA^a_{\mu}(l-l'), \nonumber
\end{eqnarray}
where $T$ means transposition;
\begin{eqnarray}
  \Gamma^{(2)}_{\bar{\psi}A} \equiv \frac{\delta^2\Gamma}{\delta_L\bar{\psi}^{a_f}(l)\delta
  A^b_{\nu}(l')}=gZ_{\psi}\sqrt{Z_A}\gamma^{\nu}t^b\psi^{a_f}(l-l'); \nonumber
\end{eqnarray}
\begin{eqnarray}
  \Gamma^{(2)}_{A\bar{\psi}} \equiv \frac{\delta^2\Gamma}{\delta A^b_{\nu}(-l')\delta_R\bar{\psi}^{a_f}(l)}=-gZ_{\psi}\sqrt{Z_A}[t^b\gamma^{\nu}\psi^{a_f}(l+l')]^T; \nonumber
\end{eqnarray}
\begin{eqnarray}
  \Gamma^{(2)}_{A\psi} &\equiv& \frac{\delta^2\Gamma}{\delta
  A^a_{\mu}(-l)\delta_R\psi^{b_f}(l')}=gZ_{\psi}\sqrt{Z_A}\bar{\psi}^{b_f}(-l+l')\gamma^{\mu}t^a; \nonumber
\end{eqnarray}
\begin{eqnarray}
  \Gamma^{(2)}_{\psi A} &\equiv& \frac{\delta^2\Gamma}{\delta_L\psi^{b_f}(l')\delta
  A^a_{\mu}(l)}=-gZ_{\psi}\sqrt{Z_A}[\bar{\psi}^{b_f}(l+l')\gamma^{\mu}t^a]^T; \nonumber
\end{eqnarray}
\begin{eqnarray}
  \Gamma^{(2)}_{AA} &\equiv& \frac{\delta^2\Gamma}{\delta A^a_{\mu}(-l)\delta A^b_{\nu}(l')}=Z_A\delta^{ab}\delta^{\mu\nu}\delta_{ll'}l^2 \nonumber\\
                    &+& gZ_A\sqrt{Z_A}f^{abc}i(2l_{\nu}-l'_{\nu})A^c_{\mu}(l-l') \nonumber\\
                    &+& gZ_A\sqrt{Z_A}f^{abc}(-i)(l_{\mu}-2l'_{\mu})A^c_{\nu}(l-l') \nonumber\\
                    &+& gZ_A\sqrt{Z_A}f^{abc}\delta^{\mu\nu}(-i)(l_{\kappa}+l'_{\kappa})A^c_{\kappa}(l-l') \nonumber\\
                    &+& g^2Z_A^2\int_q A^{c\mu}(q)A^{d\nu}(l-l'-q) \nonumber\\
                    && ~~~~~~~ \cdot(f^{eab}f^{ecd}-f^{ead}f^{ebc}) \nonumber\\
                    &+& g^2Z_A^2\int_q A^{c\lambda}(q)A^{d\lambda}(l-l'-q)\delta_{\mu\nu} \nonumber\\
                    && ~~~~~~~ \cdot\frac{1}{2}(f^{eac}f^{ebd}+f^{ead}f^{ebc}). \nonumber
\end{eqnarray}

Derivative expansions of the regulated propagators are
\begin{eqnarray}
  && \frac{1}{(\slashL{l}+\slashL{p})(1+r_{\psi}((l+p)^2/k^2))+m} \nonumber\\
  && ~\approx \frac{1}{(\slashL{l}+\slashL{p})(1+r_{\psi}(l^2/k^2))+m} \nonumber\\
  && ~= \frac{m-\slashL{l}(1+r_{\psi})}{m^2+l^2(1+r_{\psi})^2}+\frac{-\slashL{p}(1+r_{\psi})}{m^2+l^2(1+r_{\psi})^2} \nonumber\\
  && ~+ [m-\slashL{l}(1+r_{\psi})]\frac{-2p\cdot l(1+r_{\psi})^2}{[m^2+l^2(1+r_{\psi})^2]^2} \nonumber\\
  && ~+ [m-\slashL{l}(1+r_{\psi})]\frac{-p^2(1+r_{\psi})^2}{[m^2+l^2(1+r_{\psi})^2]^2} \nonumber\\
  && ~+ \slashL{p}(1+r_{\psi})\frac{2p\cdot l(1+r_{\psi})^2}{[m^2+l^2(1+r_{\psi})^2]^2} \nonumber\\
  && ~+ [m-\slashL{l}(1+r_{\psi})]\frac{4(p\cdot l)^2(1+r_{\psi})^4}{[m^2+l^2(1+r_{\psi})^2]^3}+O(p^3).
\end{eqnarray}
\begin{eqnarray}
  && \frac{1}{(l+p)^2(1+r_A((l+p)^2/k^2))} \nonumber\\
  && ~\approx \frac{1}{(l+p)^2(1+r_A(l^2/k^2))} \nonumber\\
  && ~= \frac{1}{l^2(1+r_A)}-\frac{2p\cdot l}{(l^2)^2(1+r_A)} \nonumber\\
  && ~- \frac{p^2}{(l^2)^2(1+r_A)}+\frac{4(p\cdot l)^2}{(l^2)^3(1+r_A)}+O(p^3).
\end{eqnarray}

The $\partial_t$ derivatives of the regulated propagators are
\begin{eqnarray}
  && \frac{1}{\slashL{l}(1+r_{\psi})+m}(\partial_t r_{\psi})\slashL{l}\frac{1}{\slashL{l}(1+r_{\psi})+m} \nonumber\\
  && ~ = (\partial_t r_{\psi})\frac{[m^2-l^2(1+r_{\psi})^2]\slashL{l}+2ml^2(1+r_{\psi})}{[m^2+l^2(1+r_{\psi})^2]^2} \nonumber\\
  && ~ \equiv F_1\slashL{l}+F_2 m.
\end{eqnarray}
\begin{eqnarray}
  \frac{1}{l^2(1+r_A)}(\partial_t r_A)l^2\frac{1}{l^2(1+r_A)} \equiv B.
\end{eqnarray}

\onecolumngrid \vskip 5mm

\subsection{Fermion propagator}

To get the flow equation for the fermion propagator, we encounter
\begin{eqnarray}
  && \frac{1}{2}\text{Tr}(Z_A\partial_t r_A l^2)\frac{1}{Z_A l^2(1+r_A)}g^2Z_{\psi}Z_A\bar{\psi}^{a_f}(-l+l'')\gamma^{\mu}t^a\frac{1}{\slashL{l}''(1+r_{\psi})+m}\gamma^{\nu}t^b\psi^{a_f}(l''-l')\frac{1}{Z_A l'^2(1+r_A)} \nonumber\\
  && -\frac{1}{2}\text{Tr}(Z_A\partial_t r_A l^2)\frac{1}{Z_A l^2(1+r_A)}g^2Z_{\psi}Z_A[t^b\gamma^{\nu}\psi^{a_f}(l''+l)]^T\frac{1}{\slashL{l}''^T(1+r_{\psi})+m}[\bar{\psi}^{a_f}(l'+l'')\gamma^{\mu}t^a]^T\frac{1}{Z_A l'^2(1+r_A)} \nonumber\\
  && -\text{Tr}(Z_{\psi}\partial_t r_{\psi}\slashL{l})\frac{1/Z_{\psi}}{\slashL{l}(1+r_{\psi})+m}g^2Z_{\psi}^2Z_A\gamma^{\mu}t^a\psi^{a_f}(l-l'')\frac{1}{Z_A l''^2(1+r_A)}\bar{\psi}^{b_f}(-l''+l')\gamma^{\mu}t^a\frac{1/Z_{\psi}}{\slashL{l}'(1+r_{\psi})+m} \nonumber\\
  && = g^2Z_{\psi}C_2(r)(-d)\int_p\bar{\psi}^{a_f}(p)\psi^{a_f}(p)m\cdot\int_l\bigg\{\frac{B}{m^2+l^2(1+r_{\psi})^2}+\frac{F_2}{l^2(1+r_A)}\bigg\} \nonumber\\
  && + g^2Z_{\psi}C_2(r)(d-2)\int_p\bar{\psi}^{a_f}(p)\slashL{p}\psi^{a_f}(p)\cdot\int_l\bigg\{\frac{-B(1+r_{\psi})}{m^2+l^2(1+r_{\psi})^2}+\frac{2B(1+r_{\psi})^3l^2/d}{[m^2+l^2(1+r_{\psi})^2]^2}+\frac{2F_1/d}{l^2(1+r_A)}\bigg\} \nonumber\\
  && + ~\text{higher order terms}.
\end{eqnarray}

The flow equations for $m$ and $Z_{\psi}$ can be read off as
\begin{eqnarray}
  \frac{\partial_t m}{m} &=& -\tilde{g}^2C_2(r)\bigg[d J(1,4,1;\tilde{m}^2)+d J(0,2,2;\tilde{m}^2)-\frac{(d-1)(d-2)}{d}J(1,3,1;\tilde{m}^2)\bigg], \\
  \frac{\partial_t Z_{\psi}}{Z_{\psi}} &=& -\tilde{g}^2C_2(r)\bigg[\frac{(d-1)(d-2)}{d}J(1,3,1;\tilde{m}^2)\bigg].
\end{eqnarray}

\subsection{Gluon propagator}

Here one encounters
\begin{eqnarray*}
    \text{Tr}(Z_A\partial_t r_A l^2)\frac{1}{Z_A l^2(1+r_A)}g^2Z_A^2\{...\}\frac{1}{Z_A l'^2(1+r_A)}=0.
\end{eqnarray*}
Here $g^2Z_A^2\{...\}$ indicates terms in $\Gamma^{(2)}_{AA}$
that come from the 4-gluon vertex.

To get the flow equation for the gluon propagator, we encounter
\begin{eqnarray}
  && \frac{1}{2}\text{Tr}(Z_A\partial_t r_A l^2)\frac{1}{Z_A l^2(1+r_A)}gZ_A\sqrt{Z_A}\{...\}\frac{1}{Z_A l''^2(1+r_A)}gZ_A\sqrt{Z_A}\{...\}\frac{1}{Z_A l'^2(1+r_A)} \nonumber\\
  && -\text{Tr}(-Z_c\partial_t r_A l^2)\frac{-1}{Z_c l^2(1+r_A)}gZ_c\sqrt{Z_A}f^{ace}(il_{\kappa})A^e_{\kappa}(l-l'')\frac{-1}{Z_c l''^2(1+r_A)}gZ_c\sqrt{Z_A}f^{cbe'}(il''_{\lambda})A^{e'}_{\lambda}(l''-l')\frac{-1}{Z_c l'^2(1+r_A)} \nonumber\\
  && -\text{Tr}(Z_{\psi}\partial_t r_{\psi}\slashL{l})\frac{1/Z_{\psi}}{\slashL{l}(1+r_{\psi})+m}gZ_{\psi}\sqrt{Z_A}\gamma^{\mu}t^aA^a_{\mu}(l-l'')\frac{1/Z_{\psi}}{\slashL{l}''(1+r_{\psi})+m}gZ_{\psi}\sqrt{Z_A}\gamma^{\nu}t^bA^{b}_{\nu}(l''-l')\frac{1/Z_{\psi}}{\slashL{l}'(1+r_{\psi})+m} \nonumber\\
  && =g^2Z_AC_2(G)\int_p A^e_{\mu}(-p)(\delta_{\mu\nu}p^2-p_{\mu}p_{\nu})A^e_{\nu}(p)\cdot\int_l\frac{-B}{l^2(1+r_A)}\bigg[\frac{16d-32}{d(d+2)}+\frac{d-14}{2}+\frac{8}{d}\bigg] \nonumber\\
  && -g^2Z_AN_fC(r)\int_p A^e_{\mu}(-p)(\delta_{\mu\nu}p^2-p_{\mu}p_{\nu})A^e_{\nu}(p)\cdot\int_l\bigg\{\frac{F_1(1+r_{\psi})^3l^2}{[m^2+l^2(1+r_{\psi})^2]^2}\frac{-16}{d}+\frac{F_1(1+r_{\psi})^5(l^2)^2}{[m^2+l^2(1+r_{\psi})^2]^3}\frac{64}{d(d+2)} \bigg\} \nonumber\\
  && + ~\text{higher order terms and gauge symmetry breaking terms}.
\end{eqnarray}
Here $gZ_A\sqrt{Z_A}\{...\}$ indicates terms in
$\Gamma^{(2)}_{AA}$ that come from the 3-gluon vertex.

The flow equation for $Z_A$ can be read off as
\begin{eqnarray}\label{}
    \frac{1}{2}\frac{\partial_t Z_A}{Z_A} &=& -\tilde{g}^2C_2(G)\bigg[\frac{16d-32}{d(d+2)}+\frac{d-14}{2}+\frac{8}{d}\bigg]J(2,6,0;\tilde{m}^2) \nonumber\\
                                          &-& \tilde{g}^2N_fC(r)\bigg[-\frac{8}{d}J(-1,-2,3;\tilde{m}^2)+\frac{16(d+4)}{d(d+2)}J(-2,-4,4;\tilde{m}^2)-\frac{64}{d(d+2)}J(-3,-6,5;\tilde{m}^2)\bigg].
\end{eqnarray}

\subsection{Fermion-gluon vertex}

The flow equation for the coupling can be derived from the
renormalized fermion-gluon vertex.
\begin{eqnarray}
  && \frac{1}{2}\text{Tr}(Z_A\partial_t r_A l^2)\frac{-1}{Z_A l^2(1+r_A)}g^2Z_{\psi}Z_A\bar{\psi}^{a_f}(-l+l'')\gamma^{\mu}t^a\frac{1}{\slashL{l}''(1+r_{\psi})+m}g\sqrt{Z_A}A^e_{\rho}(l''-l''')\gamma^{\rho}t^e\frac{1}{\slashL{l}'''(1+r_{\psi})+m} \nonumber\\
  && ~~~~~~~~~~~~~~~~~~~~~~~~~~~ \cdot\gamma^{\nu}t^b\psi^{a_f}(l'''-l')\frac{1}{Z_A l'^2(1+r_A)} \nonumber\\
  && -\frac{1}{2}\text{Tr}(Z_A\partial_t r_A l^2)\frac{-1}{Z_A l^2(1+r_A)}g^2Z_{\psi}Z_A[t^b\gamma^{\nu}\psi^{a_f}(l''+l)]^T\frac{1}{\slashL{l}''^T(1+r_{\psi})+m}g\sqrt{Z_A}A^e_{\rho}(-l''+l''')[\gamma^{\rho}t^e]^T\frac{1}{\slashL{l}'''^T(1+r_{\psi})+m} \nonumber\\
  && ~~~~~~~~~~~~~~~~~~~~~~~~~~~ \cdot[\bar{\psi}^{a_f}(l'+l''')\gamma^{\mu}t^a]^T\frac{1}{Z_A l'^2(1+r_A)} \nonumber\\
  && -\text{Tr}(Z_{\psi}\partial_t r_{\psi}\slashL{l})\frac{1/Z_{\psi}}{\slashL{l}(1+r_{\psi})+m}gZ_{\psi}\sqrt{Z_A}A^e_{\rho}(l-l'')\gamma^{\rho}t^e\frac{1/Z_{\psi}}{\slashL{l}''(1+r_{\psi})+m} \nonumber\\
  && ~~~~~~~~~~~ \cdot(-gZ_{\psi}\sqrt{Z_A})\gamma^{\mu}t^a\psi^{a_f}(l''-l''')\frac{1}{Z_A l'''^2(1+r_A)}(gZ_{\psi}\sqrt{Z_A})\bar{\psi}^{b_f}(-l'''+l')\gamma^{\mu}t^a\frac{1/Z_{\psi}}{\slashL{l}''(1+r_{\psi})+m} \nonumber\\
  && -\text{Tr}(Z_{\psi}\partial_t r_{\psi}\slashL{l})\frac{1/Z_{\psi}}{\slashL{l}(1+r_{\psi})+m}(-gZ_{\psi}\sqrt{Z_A})\gamma^{\mu}t^a\psi^{a_f}(l-l'')\frac{1}{Z_A l''^2(1+r_A)}(gZ_{\psi}\sqrt{Z_A})\bar{\psi}^{b_f}(-l''+l''')\gamma^{\mu}t^a \nonumber\\
  && ~~~~~~~~~~~ \cdot\frac{1/Z_{\psi}}{\slashL{l}'''(1+r_{\psi})+m}gZ_{\psi}\sqrt{Z_A}A^e_{\rho}(l'''-l')\gamma^{\rho}t^e\frac{1/Z_{\psi}}{\slashL{l}'(1+r_{\psi})+m} \nonumber\\
  && +\frac{1}{2}\text{Tr}(Z_A\partial_t r_A l^2)\frac{1}{Z_A l^2(1+r_A)}(-g^2Z_{\psi}\sqrt{Z_A})\bar{\psi}^{a_f}(-l+l'')\gamma^{\mu}t^a\frac{1}{\slashL{l}''(1+r_{\psi})+m}\gamma^{\rho}t^c\psi^{a_f}(l''-l''') \nonumber\\
  && ~~~~~~~~~~~ \cdot\frac{1}{Z_A l'''^2(1+r_A)}gZ_A\sqrt{Z_A}\{...\}^{cb,\rho\nu}_{l'''l'}\frac{1}{Z_A l'^2(1+r_A)} \nonumber\\
  && +\frac{1}{2}\text{Tr}(Z_A\partial_t r_A l^2)\frac{1}{Z_A l^2(1+r_A)}gZ_A\sqrt{Z_A}\{...\}^{ac,\mu\rho}_{ll''}\frac{1}{Z_A l''^2(1+r_A)} \nonumber\\
  && ~~~~~~~~~~~ \cdot(-g^2Z_{\psi}\sqrt{Z_A})\bar{\psi}^{a_f}(-l''+l''')\gamma^{\rho}t^c\frac{1}{\slashL{l}'''(1+r_{\psi})+m}\gamma^{\nu}t^b\psi^{a_f}(l'''-l')\frac{1}{Z_A l'^2(1+r_A)} \nonumber\\
  && -\frac{1}{2}\text{Tr}(Z_A\partial_t r_A l^2)\frac{1}{Z_A l^2(1+r_A)}(-g^2Z_{\psi}\sqrt{Z_A})[t^a\gamma^{\mu}\psi^{a_f}(l''+l)]^T\frac{1}{\slashL{l}''^T(1+r_{\psi})+m}[\bar{\psi}^{a_f}(l'''+l'')\gamma^{\rho}t^c]^T \nonumber\\
  && ~~~~~~~~~~~ \cdot\frac{1}{Z_A l'''^2(1+r_A)}gZ_A\sqrt{Z_A}\{...\}^{cb,\rho\nu}_{l'''l'}\frac{1}{Z_A l'^2(1+r_A)} \nonumber\\
  && -\frac{1}{2}\text{Tr}(Z_A\partial_t r_A l^2)\frac{1}{Z_A l^2(1+r_A)}gZ_A\sqrt{Z_A}\{...\}^{ac,\mu\rho}_{ll''}\frac{1}{Z_A l''^2(1+r_A)} \nonumber\\
  && ~~~~~~~~~~~ \cdot(-g^2Z_{\psi}\sqrt{Z_A})[t^c\gamma^{\rho}\psi^{a_f}(l'''+l'')]^T\frac{1}{\slashL{l}'''^T(1+r_{\psi})+m}[\bar{\psi}^{a_f}(l'+l''')\gamma^{\nu}t^b]^T\frac{1}{Z_A l'^2(1+r_A)} \nonumber\\
  && -\text{Tr}(Z_{\psi}\partial_t r_{\psi}\slashL{l})\frac{-1/Z_{\psi}}{\slashL{l}(1+r_{\psi})+m}(-gZ_{\psi}\sqrt{Z_A})\gamma^{\nu}t^b\psi^{a_f}(l-l'')\frac{-1}{Z_A l''^2(1+r_A)}gZ_{\psi}\sqrt{Z_A}\{...\}^{\nu\mu,ba}_{l''l'''}\frac{1}{Z_A l'''^2(1+r_A)} \nonumber\\
  && ~~~~~~~~~~~~~~~~~~~~~~~~~ \cdot gZ_{\psi}\sqrt{Z_A}\bar{\psi}^{b_f}(l'''-l')\gamma^{\mu}t^a\frac{1/Z_{\psi}}{\slashL{l}'(1+r_{\psi})+m} \nonumber\\
  && = g^3Z_{\psi}\sqrt{Z_A}\int_p\int_{p'}\bar{\psi}^{a_f}(p)\gamma^{\mu}t^aA^a_{\mu}(p-p')\psi^{a_f}(p') \nonumber\\
  && ~~ \cdot \bigg\{-\big[C_2(r)-\frac{1}{2}C_2(G)\big]\int_l B\frac{l^2(1+r_{\psi})^2(d-2)^2/d+m^2(d-2)}{[m^2+l^2(1+r_{\psi})^2]^2} \nonumber\\
  && ~~~~ -2\big[C_2(r)-\frac{1}{2}C_2(G)\big]\int_l\frac{-F_1l^2(1+r_{\psi})(d-2)^2/d+F_2m^2(d-2)}{m^2+l^2(1+r_{\psi})^2}\frac{1}{l^2(1+r_A)} \nonumber\\
  && ~~~~ -C_2(G)\int_l \frac{B}{l^2(1+r_A)}\frac{4l^2(1+r_{\psi})}{m^2+l^2(1+r_{\psi})^2}\frac{d-1}{d}-\frac{1}{2}C_2(G)\int_l \frac{1}{[l^2(1+r_A)]^2}(-4l^2F_1)\frac{d-1}{d} \bigg\} \nonumber\\
  && + ~\text{higher order terms}.
\end{eqnarray}

Then the flow equation for $gZ_{\psi}\sqrt{Z_A}$ can be read off as
\begin{eqnarray}
  \frac{\partial_t(gZ_{\psi}\sqrt{Z_A})}{gZ_{\psi}\sqrt{Z_A}} &=& -\tilde{g}^2\big[C_2(r)-\frac{1}{2}C_2(G)\big]\bigg[(d-2)J(1,4,1;\tilde{m}^2)+(d-2)J(0,2,2;\tilde{m}^2)-\frac{4(d-2)}{d}J(-1,0,3;\tilde{m}^2)\bigg] \nonumber\\
   &-& \frac{1}{2}\tilde{g}^2C_2(G)\bigg[\frac{6d-6}{d}J(1,5,1;\tilde{m}^2)+\frac{4d-4}{d}J(0,3,2;\tilde{m}^2)\bigg].
\end{eqnarray}

\vskip 5mm

\twocolumngrid

\section{Some identities}\label{App:conventions}

With the convention
$\{\gamma^{\mu},\gamma^{\nu}\}=-2\delta^{\mu\nu}I_{4\times4}$ and
$d$-dimensional Euclidian spacetime,
\begin{equation}\label{}
    \gamma^{\mu}\gamma^{\mu}=(-d)I_{4\times4}.
\end{equation}
\begin{equation}\label{}
    \gamma^{\mu}\gamma^{\rho}\gamma^{\mu}=(d-2)\gamma^{\rho}.
\end{equation}
\begin{equation}\label{}
    \gamma^{\mu}\slashL{l}\gamma^{\rho}\slashL{l}\gamma^{\mu}=(d-2)l^2\gamma^{\rho}-(d-2)2\slashL{l}l_{\rho}.
\end{equation}

\begin{equation}\label{}
    \slashL{l}\slashL{l}=(-l^2)I_{4\times4}.
\end{equation}
\begin{equation}\label{}
    \slashL{l}\gamma^{\rho}\slashL{l}=l^2\gamma^{\rho}-2\slashL{l}l_{\rho}.
\end{equation}

\begin{equation}\label{}
    \text{Tr}(\gamma^{\mu}\gamma^{\nu})=-4\delta^{\mu\nu}.
\end{equation}

Color matrix products:
\begin{equation}\label{}
    t^at^bt^a=[C_2(r)-\frac{1}{2}C_2(G)]t^b.
\end{equation}
\begin{equation}\label{}
    t^at^bf^{abc}=\frac{1}{2}iC_2(G)t^c.
\end{equation}

The following expansion will be used frequently,
\begin{equation}\label{}
    \frac{1}{A+\epsilon}=\frac{1}{A}-\frac{1}{A}\epsilon\frac{1}{A}+\frac{1}{A}\epsilon\frac{1}{A}\epsilon\frac{1}{A}-...
\end{equation}

\end{appendix}

\end{document}